\documentstyle[11pt]{article}
\def\be{\begin{equation}}
\def\ee{\end{equation}}
\def\bea{\begin{eqnarray}}
\def\eea{\end{eqnarray}}
\def\la{\label}
\def\ci{\cite}
\begin{document}

\begin{flushright}
OUTP-93-32P \\
\end{flushright}
\vspace{15mm}

 \begin{center}
 {\large\bf  Unification of couplings and soft supersymmetry breaking
terms in 4D superstring models}\\
 \vspace{7mm}
{\large  A. de la Macorra}\footnote{From 1 January 1994 E-mail:
macorra@teorica0.ifisicacu.unam.mx}
and G.~G.~Ross\footnote{SERC Senior
Fellow}\\ [5mm]
{\em Department of Physics, University of Oxford,\\
  1 Keble Rd, Oxford OX1 3NP}\\ [8mm]
 \end{center}
\vspace{2mm}

\begin{abstract}
\noindent
We consider the predictions for the hierarchy of mass scales, the
fine structure constant, the radii of compactification and the
soft SUSY breaking terms which follow if SUSY breaking is
triggered by a gaugino condensate.
\end{abstract}

\section{Introduction}
The superstring \ci{r1} offers the exciting possibility of
predicting all
the parameters of the standard model in terms of a single
parameter, the string tension. However in order to realise the
full predictive power of the superstring it is necessary to
determine the origin and effects of supersymmetry breaking. Only
after SUSY is broken are the vacuum expectation values (vevs) of
moduli determined and these determine the couplings of the
effective low energy theory. Also SUSY breaking must be
responsible for the splitting of supermultiplets allowing for the
superpartners to be heavier than their standard model partners.

In supergravity theories soft SUSY breaking terms may be added
to the lagrangian without introducing new divergences and thus
need not spoil the supersymmetric solution to the hierarchy
problem. If SUSY breaking is dominated by the F term of a
specific field the soft terms may be related in a model
independent way if supersymmetry breaking occurs in a hidden
sector with coupling to the visible sector via gravitational
effects only\cite{r34,r73,r73a}. In this letter
we will investigate the predictions for the moduli and soft terms
in superstring theories in which the origin of SUSY breaking is
through a gaugino condensate in the hidden sector.

Gaugino condensation \ci{r9a,r9} offers a very plausible
origin
for SUSY breaking for it is very reasonable to expect such a
condensate to form at a scale between the Planck scale and the
electroweak breaking scale if the hidden sector gauge group has
a (running) coupling which becomes large somewhere in this
domain. Non-perturbative studies in effective supergravity
theories resulting from orbifold compactification schemes suggest
the dynamics of the strongly coupled gauge sector is such that
the gaugino condensate will form and trigger supersymmetry
breaking.

The advantage of gaugino condensation as the underlying
supersymmetry breaking trigger is twofold. In the first place it
is generic in 4D-superstring theories to have an hidden sector
with a gauge group given by $E(6)$ or smaller. For such
asymptotically free gauge theories the coupling usually becomes
strong at low scales so gaugino condensation may be expected -
thus there is no need to postulate an additional source of
supersymmetry breaking. Secondly the scale of gaugino
condensation and the associated supersymmetry breaking scale are
dynamically determined and thus the mechanism  offers an
explanation for the magnitude of the hierarchy of masses scales.
Thus, when combined with the radiative breaking of the
electroweak symmetry, the resultant theory not only protects the
electroweak and supersymmetry breaking scales from receiving
large radiative corrections proportional to the unification or
Planck scale (the usual hierarchy problem) but also predicts the
supersymmetry breaking and electroweak breaking scales in terms
of the Planck scale and the multiplet content of the theory. In
addition, after supersymmetry breaking, the moduli of the theory
are fixed giving a prediction for the value of the gauge
couplings and the Yukawa couplings.

In order to realise the full predictive power of the 4D-
superstring theory, it is necessary to compute the non-
perturbative effects leading to gaugino condensation.  The
complete solution is clearly beyond our present-day technology
so we are forced to employ approximation methods. In this paper
we will consider the implications for the soft SUSY breaking
parameters in the case that gaugino condensation is the source
of supersymmetry breaking. We parameterise the strong binding
effects in the hidden sector by a simple effective (locally
supersymmetric) 4-gaugino coupling. We then are able to use
Nambu-Jona-Lasinio (N-J-L) techniques \cite{r10} to obtain non-
perturbative information about the gaugino binding. This shows
that a reasonable pattern of supersymmetry breaking is possible
with a single gaugino condensate. We generalize the discussion
of reference \cite{r52} to consider an orbifold model which
includes different moduli fields and we determine the tree level
potential and the one-loop radiative corrections due to strong
gaugino binding. We show that minimization of the potential leads
to vacuum expectation values (v.e.v.s) for the moduli which
either take dual invariant values or have a common value related
to the v.e.v. of the dilaton. This  naturally allows for squeezed
orbifolds which were found to be better candidates for minimal
string unification \ci{r34}. We also include chiral
matter fields in the visible sector and determine the soft SUSY
breaking terms. Unlike results obtained previously we find that
the radiative
corrections stabilise the potential for vanishing v.e.v.s of the
matter fields. We also find values for the soft parameters which
lie outside the favoured range of ``model'' independent
parameterisations of the soft terms due to our inclusion of large
non-perturbative effects beyond string tree level in the
effective potential. Of course, due to the simplicity of our
model for strong binding, our results should be considered as
only indicative of those to be expected in a complete gauge
theory. However, they do show the importance of including the
full nonperturbative effects of gaugino binding before extracting
predictions for SUSY breaking phenomena.

We start with the effective theory following from a 4-D
superstring orbifold model. It is specified by the Kahler
potential $G=K+ln\frac{1}{4}|W|^{2}$ and the gauge kinetic
function $f$ given by
\be
K=-ln(S+\bar{S}+2\Sigma_{i}k_{a}\delta_{GS}^{i}lnT_{ri})-
\Sigma_{i}ln(T+\bar{T})_{i} +
\Sigma_i K^{i}_{i}(T,\bar{T})|\varphi_{i}|^{2},
\la{a7}
\ee
\be
W=W_{0}+W_{m},
\la{eq:W}\ee
and
\be
f_{0}=S+2\Sigma_{i}(b'^{i}_{a}-k_{a}\delta^{i}_{GS})ln\,(\eta(
T_{i})^{2})
\la{a8}\ee
where
\begin{equation}
b'^{i}_{a}=\frac{1}{16\pi^{2}}(C(G_{a})-
\Sigma_{R_{a}}h_{R_{a}}T(R_{a})(1+2n^{i}_{R_{a}}))
\label{eq:gg}
\end{equation}
where $C(G_{a})$ is the quadratic Casimir operator of the gauge
group
$G_{a}$,
$k_{a}$ is the level of the corresponding   Kac-Moody lalgebra
for the
gauge group $G_{a}$ and
$\delta_{GS}^{i}$ is the Green-Schwarz term needed to cancel the
gauge
independent part of the duality anomaly \ci{r34,r19,r18}. $W_{0}$
in eq.(\ref{eq:W}) is the scalar potential due to the gaugino
condensate (see below) and  $W_{m}$ the matter superpotential.
The subscript, 0, on the gauge kinetic function $f$ indicates
that it is given at the string scale and the form of
eq.(\ref{a8})
includes the contribution from the massive Kaluza-Klein modes.

The normalization of the kinetic term for the chiral matter
superfields is given by $K_{i}^{i}(T,\bar{T})$ which is a
polynomial function of the moduli fields \ci{r36},
\be
K^{i}_{i}=\Pi_{j}\,\,a_{i}\,T_{Rj}^{n_{ij}}
\la{a9}\ee
where $a_{i}$ a constant (in most cases $a_{i}=1$) and $n_{ij}$
is the modular weight of the superfield $\varphi_{i}$ with
respect to the real part, $T_{Rj}=T_{j}+\bar{T}_{j}$, of the
moduli $T_{j}$.
Eq.(\ref{a7}) was derived  for
$K_{i}^{i}(T,\bar{T})|\varphi_{i}|^{2} \ll 1$.
Note that the Kahler potential includes the Green-Schwarz term
and the dilaton transforms under duality in a non-trivial way
($S\rightarrow S +
2\Sigma_{i}k_{a}\delta^{i}_{GS}ln(icT_{i}+d)$) so that
$Y\equiv S+\bar{S}+2\Sigma_{i}k_{a}\delta_{GS}^{i}lnT_{ri}$ is
duality invariant.

 The form of this effective theory is fixed by the requirement
that it should correspond to the strongly coupled theory in the
hidden sector. We have suggested a simple approximation to this
structure which, we hope, captures the essentials of the strong
binding effects leading to gaugino condensate. The basis of this
approximation is to start with a (N=1 locally supersymmetric)
four fermion description of the strong hidden sector interaction
and to use Nambu-Jona-Lasinio techniques to compute the non-
perturbative effects responsible for gaugino condensation. The
model is defined by \ci{r52,r15}
\be
W_{0}=\Pi_{i}\eta^{-2}(T_{i})\,\,\Phi
\la{a11}
\ee
\be
f=f_{0}+\xi ln(\Phi/\mu)
\la{a12}
\ee
where $\Phi$ is an auxiliary chiral superfield and
$\xi=\frac{2b_{0}}{3}$.
Duality invariance fixes the mass parameter $\mu$ in
eq.(\ref{a12}) to
be $\mu=m\Pi_{i}T_{ri}^{-3b'_{ai}/2b_{a}}$ with $m$ duality
invariant.
We choose $m=M_{s}^{3}$  so that the parameter $\mu$ becomes,
at the field theoretical limit,
$\mu=M^{3}_{com}$, with $M_{com}=M_{s}\Pi_{i}T_{ri}^{-1/6}$ the
compactification scale.
Since $\mu$ is
non-holomorphic, consistency with supersymmetry requires to
interpret $\mu$ as wave function renormalization of the dilaton
field. In such a case $\mu$ would appear in the Kahler potential
and not in the kinetic function \ci{r19}.

 From the classical equation of motion the scalar component of the
auxiliary field $\Phi$ is given in terms of the gaugino bilinear
by \ci{r52}
\be
\phi=\frac{e^{-K/2}\xi}{2m^{2}\Pi_{i}\eta^{-2}_{i}}\,
\bar{\lambda}_{R} \lambda_{L}
\la{a13}
\ee
while the fermion component is given in terms of the gaugino and
chiral fermions.

For a pure gauge theory in the hidden sector the tree level
scalar potential is  \cite{r15}
\be
V_{0}=  h_{i}(G^{-1})^{j}_{i}h^{j}-3e^{G} \la{a15}\ee
with $i=S,T$ and $h_{i}$  the F-term of the chiral
superfields $S$ or $T_{i}$.

For the choice of $W$ and $f$ given
in eq.(\ref{a11}) and eq.(\ref{a12}) the F-terms
for the $S$ and $T_{i}$ fields are
\be
h_{S}=-e^{G/2}G_{S}+\frac{1}{4} f_{S}
\bar{\lambda}_{L}\lambda_{R}=\frac{1}{2}e^{K/2}W_{0}\frac{(1+Y
/\xi)}{Y} \la{a18}\ee
and
\be
h_{T_{i}}=\frac{1}{2}e^{K}W_{0}\left[\frac{1}{T_{ri}}(1+\frac{
a_{i}}{Y})+2\frac{\eta_{T_{i}}}{\eta}(1-\frac{a_{i}}{\xi})\right]
\la{ac14}\ee
where $a_{i}=2(k_{a}\delta_{i}^{GS}-b_{ai})$.
The tree level scalar potential is then
\be
V_{0}=m^{2}_{3/2} B_{0}
\la{a20}\ee
with
\be
B_{0}=\left[(1+\frac{Y}{\xi})^{2}+\Sigma_{i}\frac{Y}{Y+a_{i}}(
1-\frac{a_{i}}{\xi})^{2}\,\frac{T_{ri}^{2}}{4\pi^{2}}|\hat{G}_
{2}|^{2}-3\right]
\la{a21}
\ee
and $\hat{G}_{2}$ the Eisenstein modular form with weight 1/2.
The gravitino mass is given by
\be
m^{2}_{3/2}=\frac{1}{4}e^{K}\Pi_{i}|\eta(T_{i})|^{-4}\,\,|\phi
|^{2}.
\la{a22}\ee
Clearly SUSY will be broken and a non-zero  gravitino mass will
result only if there is a non-vanishing  v.e.v for $\phi$
generating a non-zero $h_{S}$ (cf. eq.(\ref{a18})). Assuming this
to be the case and solving eq.(\ref{a12}) gives a
parameterisation of gaugino condensation equivalent to that
derived by other methods \cite{r50}-\ci{r55}.  To this extent the
form of
eq.(\ref{a20}) is just a re-parameterisation of the usual model
of
gaugino condensation. Where we depart from this analysis is our
inclusion of radiative corrections in the effective potential.
Summing these effects to all orders shows that the gaugino
condensate dynamically forms.

Using eq.(\ref{a13}) the scalar potential may be written in terms
of the gaugino field
\be
V_{0}=\frac{\xi^{2}H}{16(Ref)^{2}}\,|\bar{\lambda}_{R}'\lambda
_{L}'|^{2} \la{e29}\ee
where the factor of $(Ref)^{2}$ in the denominator in
eq.(\ref{e29}) appears because we have rescaled the gaugino
fields appearing in this equation to have canonical kinetic
terms. Thus we see that the choice of $W$ and $f$ in
eqs.(\ref{a11}) and (\ref{a12}) leads to a four-fermion
interaction with strength proportional to $f^{-1}$.

In a pure gauge theory the running gauge coupling is related to
$f$ by $g_s^2=(Re f)^{-1}$ with
\be \frac{1}{g_{a}^{2}(\Lambda)}=\frac{k_{a}}{g^{2}_{s}}
+b_{a}ln(\frac{\Lambda^{2}}{M_{a}^{2}})+\Delta_{a}
\la{a1}
\ee
where $k_a$ is the level of the corresponding Kac-Moody algebra
for the gauge group $G_a$, $b_{a}=\frac{1}{16\pi^{2}}(3C(G_{a})-
\Sigma_{R_{a}}h_{R_{a}}T(R_{a}))$ is the N=1 $\beta$-function
coefficient and $h_{R_{a}}$ the number of chiral fields in a
representation $R_{a}$.  $M_{a}$ is the renormalization scale
below which the coupling constants begins to run
\be
M^{2}_{a}=\Sigma_{i}\,(T_{ri})^{\alpha_{ai}}\,M_{s}^{2}
\la{a2}
\ee
where $T_{Ri}$ and
the constants $\alpha_{ai}$ is model and gauge dependent. \be
\alpha_{ai}=\frac{k_a\delta_{GS}^i-b_a^{\prime i}}{b_a}
\la{alpha}
\ee
In the case of a single  overall (1,1) moduli $T_{i}=T$,
$i=1,2,3$,   with $\alpha=\Sigma\alpha_{i}=-1$ one obtains the
``naive'' field theoretical expression for $M_{a}=(Re S Re
T)^{-1/2}$. The existence of an infinite number
of massive states (Kaluza and winding states), above the string
scale, give rise to string threshold contribution $\Delta_{a}$
\ci{r17,r17a} which are relevant to the determination of the
coupling constant at the string scale (cf. eq.(\ref{a8})). In the
effective four fermion theory defined by eqs.(\ref{a11}) to
(\ref{e29}) $f^{-1}$ sets the scale for the strength of the
interaction (cf. eq.(\ref{e29})). If the theory is to model the
strong binding effects due to a gauge interaction it clearly
should be related to the gauge coupling  $g_s^2$.
Thus $f^{-1}$ does indeed run like $g^2$.
However this running gives rise to an unphysical singularity in
$f^{-1}$ related to the Landau singularity which we will choose
to regulate when applying eq.(\ref{a13}). Below the scale at
which the running coupling becomes large the hidden sector states
are confined and have a mass of $O(\Lambda_c$). Thus we expect
the 4 Fermion effective interaction to be pointlike at scales
below $\Lambda_c$ and so we cutoff the growth implied by
eq.(\ref{e29}) when the strength of the interaction reaches
$1/\Lambda_c^2$.

The tree level potential for $\phi$, given by
eqs.(\ref{a20}),(\ref{a21}) and (\ref{a22}),
has no stable solution.
This is consistent with the observation \cite{clmr}
that gaugino condensation as usually parameterised does not occur
in models with a single hidden sector gauge group factor. However
we have argued it is essential to go beyond tree level to include
non-perturbative effects in the effective potential which may
allow for a non-trivial minimum even in the simple case of a
single hidden sector gauge group. This non-perturbative sum
(equivalent to the NJL sum) is readily obtained simply by
computing the one loop correction to V. If these destabilise the
potential the resultant minimum will correspond to a cancelation
of tree level and one-loop  terms which, as noted above, is
necessarily non-perturbative in character \cite{r10}.

The one-loop radiative corrections may be calculated using the
Coleman-Weinberg one-loop effective potential \ci{r23,r57}, \be
V_{1}=\frac{1}{32\pi^{2}} Str \int d^{2} p\, p^{2}
ln(p^{2}+M^{2}) \la{a24}\ee
where $M^{2}$  represents the square mass matrices and $Str$ the
supertrace.
As discussed above the 4 Fermi interaction has a form factor
which falls rapidly above $\Lambda_c$ effectively giving a
momentum space cutoff at the condensation scale
defined as the scale where the gauge coupling constant (cf.
eq.(\ref{a1})) becomes strong.

The leading contribution to the one-loop potential
$V_{1}$ comes from  the supersymmetry breaking mass of the
gaugino fields and keeping only these terms one has
\be
V_{1}=-\gamma \Lambda_{c}^{4}J(x_{g})
\la{a29}\ee
with
\be
J( x)=x+x^{2} ln(\frac{x}{1+x})+ln(1+x)
\la{a26}\ee
$\gamma=\frac{n_{g}}{32\pi^2}$ ($n_{g}$ the dimension of the
hidden gauge group) and
$x_{g}=\frac{m^{2}_{g}}{\Lambda^{2}_{c}}$. The tree level gaugino
mass is given by
\begin{eqnarray}
m_{g}& = & \frac{2}{Ref}\frac{\partial V_{0}}{\partial
W_{0}}\frac{\partial^{2} \bar{W}_{0}}{\partial \bar{\lambda}_{R}
\partial \lambda_{L}} \nonumber \\
 & = & \frac{b_{0}}{3Ref}\,B_{0}m_{3/2}
\la{a30}
\end{eqnarray}
with $B_{0}$ given in eq.(\ref{a21})
and  the factor $Ref$ in eq.(\ref{a30}) is due to non-canonical
kinetic terms for the gauginos. In addition there should be  a
supersymmetric contribution to the mass of the hidden sector
states, gauginos and gauge bosons, generated by the strong hidden
sector forces which  (in analogy with QCD) may be expected to be
confining. Since the inclusion of these supersymmetric masses
only results in a shift of the v.e.v. of the dilaton field we
will not consider them in this paper.

We now consider the minimisation of the full effective potential,
$V_0+V_1$. As discussed in \cite{r52,r56}, a supersymmetry
breaking solution with non zero $\phi$ is found for a large four
fermion coupling. To regulate the unphysical divergence coming
from the vanishing of $Re \; f$ we cut off its evolution at $Re
\; f=\Lambda_c$, the point at which the gauge coupling becomes
large. This is done to give the dimensionally correct form for
the effective four fermion interaction at low energies,
$\frac{1}{\Lambda_c^2}(\bar{\psi}\psi)^2$. With this there is a
well defined minimum with the condition $\frac{\partial
V}{\partial \phi}=0$ being satisfied through a cancelation of
tree level with one loop terms. As such it is necessarily non-
perturbative and shows that, within our four fermion
approximation, gaugino condensation is dynamically generated. The
condition $\partial \; V/\partial \; T_i$ can be satisfied either
for $T_{i}$ at the dual invariant points
($T_{i}=1,e^{i\pi/6}$) or for large values of $T_i$ corresponding
to squeezed orbifolds. In the latter case only if
$\alpha_{ai}=\alpha_{aj}=\alpha$ it is possible that different
moduli $T_{i}$ and $T_{j}$ have a large v.e.v and their v.e.v.s
will necessarily be the same (up to modular invariant
transformations), i.e. $T_{i}=T_{j}$. We see that it is thus
only possible to have a large overall moduli $T$ if all three
$\alpha_{i}$ have the same value, as in the case for a $Z_{3}$
or $Z_{7}$ orbifold where all $\alpha_{i}$ vanish \ci{r18}. For
an orbifold with two completely rotated  planes then only one
$\alpha_{i}$ is (possibly) different than zero in which case the
v.e.v. for the moduli attached to this plane  will be different
than those of the other two.
A squeezed orbifold may thus naturally be obtained\footnote{These
orbifolds were found to be better candidates for a
minimal string unification \cite{r56},\cite{r34}.}.
The v.e.v. of $Y$ and $x_g$ are given in terms of the dimension
of the hidden gauge group, its  one-loop N=1 $\beta$-function
coefficient and  $\alpha$ by \ci{r56}
\be
Y=8\pi\sqrt{\frac{1}{n_{g}}(1+\frac{2\alpha-1}{3\alpha-1}\,
\epsilon)^{-1}}
\la{a36}
\ee
and
\be
\epsilon\equiv x_gln(x_g)=\frac{4b_{0}}{Y}(3\alpha-1)
\la{eq:e}
\ee
 From eq.(\ref{eq:e}) we obtain the value of the moduli
\be
\Sigma_{i}(1-\alpha_{i})T_{ri}=\frac{3Y}{\pi b_{0}}-
\frac{6}{\pi}ln(Q)
\la{a37}\ee
where the sum in eq.(\ref{a37}) is over all moduli that acquire
a v.e.v different from the dual invariant points  and
$Q\equiv\frac{9Y^{3}}{64b_{0}^{2}}\frac{\Pi_{i}T_{ri}^{(\alpha
-1)}M_{s}^{2}}{x_{g}}$.  If the gauge group is broken down from
$E_{8}$ to a lower rank group such as  $SU(N)$ with $5\le N \le
9$  (as can be easily done by compactifying on an orbifold with
Wilson lines)  a phenomenologically interesting
solution can be obtained with  only one gaugino
condensate. The value of the dilaton is mainly given by the
dimension of the hidden gauge group  and for a phenomenologically
interesting solution one needs
 $\frac{Y}{b_{0}(1-\alpha)}\simeq 58$.
In particular if the hidden sector gauge group is $SU(6)$ the
gauge coupling constant at the unification scale would be
$g^{-2}_{s}=Y/2\simeq 2.1$.  Using eq.(\ref{a36}),
the fine structure constant at
the unification scale is given by
\be
\hat{\alpha}^{-1}_{gut}=(\frac{g^{2}_{s}}{4\pi})^{-1}\simeq
\frac{16\pi^{2}}{\sqrt{ng}}.
\la{eq:sc}\ee
For an $SU(6)$ gauge group $\alpha^{-1}_{gut}\simeq 26.7$. This
value is consistent with minimal
supersymmetric standard model (MSSM) unification \ci{r58}.

On the other hand, the v.e.v. of the moduli fields does
depend  on $\alpha_{ai}$, the v.e.v. of the dilaton and $b_{0}$.
The value of $\alpha$ is approximately $-1/3$  if either the
modular weight of the hidden matter fields is
$n^{ij}_{R_{a}}=-1/3$ (i.e. untwisted fields) or if the
dominant term in the N=1 $\beta$-function coefficient for the
hidden gauge group
$b_{0}$  is given by the quadratic Casimir operator $C(G_{a})$,
 in which case from eq.(\ref{alpha})  (taking
$\delta_{GS}=0$)
 \[ \alpha_{i}=-\frac{b'_{0i}}{b_{0i}}=-
\frac{1}{3}(1-
\frac{2}{C}\Sigma_{R_{a}}h_{R_{a}}T(R_{a})(\frac{1}{3}+n^{i}_{
R_{a}}))\simeq-\frac{1}{3}
\]
If we only consider the three diagonal
(1,1) moduli (which are always present in orbifold
compactification) then, for a  gravitino mass of order $1\,TeV$
and $\alpha=-1/3$, $ReT\simeq 8,12,22$ depending on whether
three, two or one moduli get a ``large'' v.e.v., respectively.
In the case discussed above of a large  overall modulus
$ReT=ReT_{1}=ReT_{2}=ReT_{3}\simeq 8$. Note that this result is
quite different from the values previously obtained $T\simeq 1.3$
\ci{r26}. Clearly the discrepancy is due to the inclusion of the
strong gaugino binding effects in $V_{1}$ (cf. eq.(\ref{a29})).

Interesting enough these large values
are necessary for the consistent minimal unification
scenario with the correct values of the weak angle
$sin\theta_{w}$ and
strong coupling constant $\alpha_{strong}$ \ci{r54}.

Up to now we have only considered the scalar potential for
gauginos in the hidden sector, dilaton and  moduli
 fields. We would now like  to include  the contribution from
matter superfields.
In general the matter superpotential is
given as a power series expansion in the chiral superfields
$\varphi$ suppressed  by the Planck mass. In the absence of
linear and quadratic terms, the leading term in the
superpotential is cubic in the matter superfields since one
assumes that the matter fields at most get a  small v.e.v.
compared to the Planck mass to be consistent with the Kahler
expansion in eq.(\ref{a7}), i.e.
$|\varphi_{i}|^{2}K_{i}^{i}(T,\bar{T})\le  1$.
The gaugino condensate term $W_{0}$ will then
give  the dominant contribution to the scalar potential and the
v.e.v. of the dilaton and moduli fields
obtained in eqs.(\ref{a36}) and (\ref{a37})  will remain valid.

For a generic matter superpotential  $W_{m}$, using the Kahler
potential of eq.(\ref{a7}), the tree level scalar potential is
\ci{r56}
\be
V_{0}=\frac{1}{4}e^{K}\left((K^{-1})^{S}_{S}|F_{S}W_{0} +K_{S}
W_{m}|^{2} + \Sigma_{i}(K^{-1})^{i}_{i}|F_{i}W_{0}+\beta_{i}|^{2}
- 3|W_{0}+W_{m}|^{2}\right) \la{a45}\ee
where
\[
F_{S}=K_{S}-\frac{3}{2b_{0}}f_{S}=-
\frac{1}{Y}(1+\frac{3Y}{2b_{0}}), \]
\[F_{i}=K_{i}+\frac{W_{0,i}}{W_{0}}-\frac{3}{2b_{0}}f_{i}\]
\[\beta_{i}=K_{i}W_{m}+W_{m,i}, \]
$b_{0}$ is the one-loop beta function coefficient for the hidden
gauge group, $f_{i}$ is the derivative of the gauge kinetic
function with respect to the i-field   and the index $i$ runs
over all matter and moduli fields. The first term in
eq.(\ref{a45}) is just the square of the auxiliary field of the
dilaton $h_{S}$ while the second term  is a sum of the squares
of the auxiliary fields of the moduli $h_{Ti}$ and chiral matter
fields $h_{\varphi_{i}}$.

To determine the vacuum structure one has to
include quantum corrections and  the main contribution to the
one-loop potential is again given by the gaugino-loops.
 We would like to emphasise again that, although we are just
calculating the one-loop potential, the final result is
necessarily non-perturbative, since by solving the mass gap
equation one is stabilizing the tree level potential by the one-
loop potential, and one is effectively summing an infinite number
of gaugino-bubbles.  One can easily determine  the contribution
of these loops by calculating  the gaugino mass at tree level and
then using eq.(\ref{a29}). Since all dependence on the gaugino
bilinear in eq.(\ref{a13}) is given in $W_{0}$ one has,
\be
m_{g}=\frac{b_{0}}{6Ref}e^{K/2}\,H
\la{a46}\ee
with
\be
H\equiv F^{S}(K^{-1})^{S}_{S}(F_{S}
W_{0}+K_{S}W_{m})+F^{i}(K^{-1})^{i}_{i}(F_{i}
W_{0}+\beta_{i})-3(W_{0}+W_{m}). \la{a47}\ee
The one-loop potential (considering  the contribution from the
gauginos-loops only) is  (cf. eq.(\ref{a29}))
\[
V_{1}=-\gamma \Lambda_{c}^{4}J(x_{g})
\]
\be
V_{1}=-\gamma \Lambda_{c}^{4}(2x_{g}+x^{2}_{g}ln(x_{g}))
\la{a48}\ee
where we have expanded the function $J$  since, as we showed  in
eq.(\ref{eq:e}), $x_{g}=O(10^{-2})$  and  it is now given by  \be
x_{g}  =  \frac{m^{2}_{g}}{\Lambda_{c}^{2}}=\frac{b_{0}^{2}}{36}
\frac{e^{K}}{\Lambda_{c}^{4}}\,|H|^{2}. \la{a49}\ee
If  $W_{m}=W_{m,i}=0$ one has $H=B_{0}W_{0}$ and eq.(\ref{a46})
reduces to eq.(\ref{a30}), as should be the case.
The scalar potential $V=V_{0}+V_{1}$ is then
\begin{eqnarray}
V & = & \frac{1}{4}e^{K}[(K^{-1})^{S}_{S}|F_{S}W_{0} +
K_{S}W_{m}|^{2} +
\Sigma_{i}(K^{-1})^{i}_{i}|F_{i}W_{0}+\beta_{i}|^{2}
\nonumber \\
 & & -3|W_{0}+W_{m}|^{2}
-\frac{2\gamma b_{0}^{2}}{9}|H|^{2}
(1+\frac{1}{2}x_{g}ln(x_{g}))]
\la{a50}
\end{eqnarray}
 So far we have not discussed the cosmological constant at the
minimum of the potential. In fact, in the absence of matter
fields, it is negative, of order $\Lambda_c^4$. It may be
cancelled through the addition to the matter superpotential of
a term linear in a chiral superfield D \ci{r56}
\begin{equation}
W_m=c \; D
\label{eq:cc}
\end{equation}

To obtain a more transparent expression for $V$
one can expand the scalar potential in powers of the gravitino
mass ($m^{2}_{3/2}=\frac{1}{4}e^{K}|W_{0}|^{2}$). Making use of
the solutions to the extremum equations eq.(\ref{a36}) and
eq.(\ref{a37}) the scalar potential is
\be
V=m^{2}_{3/2}B +\, (m_{3/2}A\,+\,h.c.) + \,C
\la{a51}\ee
with
\[B=\frac{1}{2}((K^{-1})^{S}_{S}F^{S}F_{S}+(K^{-1})_{i}^{i}K_{i}
K^{i}+K_{D}^{D}|D|^{2}-3)(\frac{\alpha-1}{3\alpha-1}-
K_{D}^{D}|D|^{2})\,\epsilon, \]
\be
A=\frac{1}{4}e^{K/2}
\left(D(K_{D}^{D}|D|^{2}c+c)+F^{S}(K^{-1})_{S}^{S}K_{S}
cD-3cD\right)(\frac{\alpha-1}{3\alpha-1}-
K_{D}^{D}|D|^{2})\,\epsilon
\la{a59}
\ee
and
\[C=\frac{1}{4}e^{K}((K^{-1})_{i}^{i}|K_{i}W_{m}+W_{m,i}|^{2}
+(K_{D}^{D})^{-1}|K_{D}^{D}|D|^{2}c+c|^{2} -3|cD|^{2})\]
where the $i$-index in eqs.(\ref{a59}) runs over all chiral
superfields but for  $D$.
The leading contribution from the superpotential of
eq.(\ref{eq:cc}) is through the auxiliary field of $D$
($h_{D}=\frac{1}{2}e^{K/2}(K_{D}W+W_{D})$).  This contribution
is semipositive definite and for vanishing v.e.v. of $D$ the only
term that survives is the one proportional to
$|W_{m,D}|^{2}=c^{2}$ and the
one-loop potential is independent of it. It is in
fact this term  that gives the  main contribution to the scalar
potential even for non-vanishing v.e.v. of $D$ and since it is
positive it allows for the cancellation of the cosmological
constant by fine tuning $c$. While such a tuning is unnatural,
it does demonstrate that it is possible to have a zero
cosmological constant with supersymmetry broken at a
phenomenologically realistic value.

  It is now straightforward to compute the soft supersymmetry
breaking masses in the visible sector. The main contribution
comes from the auxiliary field of the dilaton $h_{S}$ as can be
seen from eqs.(\ref{a51}) and (\ref{a59}) since the v.e.v. of
$h_{S}^{2}\gg h_{Ti}^{2}$. The leading contribution to the scalar
mass $m_{0}$ comes from the $e^K$ dependence implicit in B and
gives (assuming zero cosmological constant)
\be
m_{0}^{2}=-\frac{1}{2}\,B_{0}\,\epsilon
\left(\frac{3-5\alpha}{(3\alpha-2)(\alpha-3)}\right)^2
\,m^{2}_{3/2}
\la{a57}
\ee
Since $\epsilon<0$ and $\alpha< 1/3$
eq.(\ref{a57}) is positive showing, contrary to analyses
neglecting the effect of $V_1$, that the scalar potential has a
minimum for $\phi_i=0$. Note that the scalar mass is the same for
all matter superfields regardless of their modular weights. Since
the auxiliary field of the dilaton is much bigger then the
auxiliary field of the $D$ field  the scalar mass is  only
sensitive to the fact that one has no cosmological constant and
not to the details of the cancelation of the vacuum energy.  In
the absence of v.e.v.s for the matter fields, the
common supersymmetric mass is independent of the trilinear
superpotential.

Using eqs.(\ref{a15}), (\ref{a18}) and (\ref{a24}) one obtains
the leading contribution to the common gaugino mass, $m_{1/2}$,
of the observable sector
\be
m_{1/2}=-6\alpha\left(\frac{3-5\alpha}{(3\alpha-2)(\alpha-3)}
\right)\,m_{3/2}
\la{a63}
\ee
This  comes from the $h_s$ auxiliary field contribution and is
the same for all gaugino masses at the
unification scale where  the gauge coupling constants meet.

The remaining soft terms are proportional to the A component of
the matter superpotental, $W_m \mid_A$. They are given by
\begin{equation}
L=\Sigma_NR_N \; m_{3/2}\tilde{W}_m^N + h.c.
\label{eq:asoft}
\end{equation}
where N is the degree of the normalized superpotential
$\tilde{W}_m^N=\frac{1}{2}e^{K/2}W_{m}^{N}$  and
\begin{equation}
R=\frac{3(\alpha-1)}{(3\alpha-2)(\alpha-3)}\left[(3-5\alpha)
+\frac{(\alpha-1)}{2}(1+N+K^{T}(K^{-1})^{T}_{T}\,
\frac{h_{123,T}}{h_{123}})\right].
\label{eq:R}
\end{equation}
For the leading trilinear terms,
$\tilde{W}^3_m=\frac{1}{2}e^{K/2}h_{ijk} \phi_i\phi_j \phi_k$,
whose coupling, $h_{ijk}$, is independent of
the moduli $T$ this reduces to
\begin{equation}
A_{t}\equiv
R_3=\frac{3(\alpha-1)(1-3\alpha)}{(3\alpha-2)(\alpha-3)}.
\label{eq:R3}
\end{equation}
For a bilinear term, such as the MSSM term $\tilde{W}^2_m=\mu H_1
H_2$, with $\mu$ independent of $T$ we have
\begin{equation}
B'\equiv
R_2=\frac{3(\alpha-1)(3-7\alpha)}{2(3\alpha-2)(\alpha-3)}.
\label{eq:R2}
\end{equation}
One problem associated with such a term is that there is no
obvious reason why $\mu\le O(M_W)$ as must be the case if the
hierarchy problem is not to be re-introduced. To answer this it
has been suggested that the $\mu$ term originates from a term in
the Kahler potential proportional to $H_1H_2$ \ci{r67}. After
supersymmetry breaking this generates the $\mu H_1 H_2$ term in
the superpotential with $\mu \propto m_{3/2}$. In this case the
soft term coming from the B term in eq.(\ref{a51}) is again of
the form of eq.(\ref{eq:asoft}) with
\bea
B''&=&\left(\frac{m_{0}}{m_{3/2}}\right)^{2}
\label{eq:R2P}\\
&=&\left(\frac{3-5\alpha}{(3\alpha-2)(\alpha-3)}\right)^{2}
\frac{\epsilon
B_{0}}{2}
\eea
Another suggestion is that the $\mu$ term comes from a term in
the superpotential proportional to $W_0H_1H_2$. In this case we
find from the B term the contribution of the form of
eq.(\ref{eq:asoft}) with
\bea
B'''&=&\frac{(\alpha-1)}{(3\alpha-1)}\left(\frac{m_{0}}{m_{3/2
}}\right)^{2}
\nonumber\\
&=&\frac{(\alpha-1)}{(3\alpha-1)}\left(\frac{3-5\alpha}
{(3\alpha-2)(\alpha-3)}\right)^{2}\frac{\epsilon
B_{0}}{2}
\label{eq:R2PP}
\eea

This completes the derivation of the soft terms arising from
gaugino condensation. It is also interesting to calculate  the
fine tuning parameter $\Delta$ \ci{r65}, defined as
\be
|\frac{a_i}{M^{2}_{Z}}\frac{\partial M^{2}_{Z}(a_i,h_i)}{\partial
a_i}| < \Delta
\la{eq:D}\ee
where $a_{i}$ are the soft SUSY parameters and $M_{Z}$ the mass
of $Z$ boson.
This quantity measures the allowed degree of
the fine tunning needed to get the correct mass $M_Z$; for
$\Delta=10$ the value of $a_{i}$ must be fine tuned to $1/10$ of
the $Z$ mass. At a crude approximation
$\Delta\simeq m_{1}^{2}/M^{2}_{Z}$ with $m_{1}$
evaluated at the electroweak scale
and it is given by
$m_{1}^{2}=m_{0}^{2}+\mu^{2}+km_{1/2}^{2}$ with $k\simeq 1/2$.

It is illuminating to compute them for
realistic examples with an
$SU(6)$ gauge group in the hidden sector with  different matter
content.
They are shown in {\bf tables 1-3}. Notice that the fine
structure constant at the unification scale
is of the right order of magnitude as required by MSSM \ci{r58}.
It is the dimension of the gauge group in the hidden sector  that
sets the value of $\hat\alpha_{gut}$(cf. eq.(\ref{eq:sc})).

These results suggest that $m_0/m_{1/2}$ is large contrary to the
analyses based on assuming supersymmetry breaking originates at
tree level from the dilaton F term even though in this case too
the dilaton F term dominates. There are two reasons for this. In
the first place the mechanism explored here, being
nonperturbative, necessarily requires that loop corrections to
the potential be as important as the tree level contribution.
Indeed there is a strong  suppression of the $R_N$ terms due to
a cancelation of the tree and loop contributions. Secondly the
cancelation of the cosmological constant in gaugino condensation
requires a contribution from the matter superfield sector. Both
these effects are absent in \cite{r73} and explains the
difference between the resulting soft terms.

For a wide range of  values for $\alpha$, a large hierarchy can
be obtained.
However, the soft supersymmetric terms are more sensitive  to
$\alpha$.  In {\bf fig.1} we
show $m_{0}/m_{3/2}$, $m_{1/2}/m_{3/2}$ and the trilinear term
$A_{t}$ as a function of $\alpha$.
For increasing $\alpha$,\, $m_{0}$ and  $A_{t}$ become
smaller. The common supersymmetric mass,  $m_{0}$,  is always
larger than
$m_{3/2}$.  On the other hand, the gaugino mass is always smaller
than
$m_{3/2}$ and it approximates  the gravitino mass for
$|\alpha|\rightarrow 1/3$. The ratio of $m_{1/2}/m_{3/2}$ is
independent
of the value of $m_{3/2}$, but $m_{0}/m_{3/2}$ is not. The reason
is
that   $m_{0}/m_{3/2}$ depends on $Y/b_{0}$, which sets the
hierarchy.

Running the masses down to the
electroweak scale, neglecting corrections due to possible large
Yukawa couplings, the common supersymmetric mass remains the same
while  the gluino mass increases by a factor
of three \ci{r63}.  If we demand all supersymmetric masses to be
in the range $100\,TeV\le m_{ss}\le 1000\,TeV$, then
$m_{0}/m_{1/2}|_{gut}\le 30$.

{}From the range of values of the parameter
$\alpha$, the preferred value is for $\alpha$ positive and close
to $1/3$. For this value of $\alpha$ the difference between the
supersymmetric masses of the scalar and  gaugino fields is
smallest. This allows for having a small fine tuning parameter
$\Delta$.
At the same time, the trilinear term $|A_{t}|$ decreases. This
is welcome since one-loop diagrams,  associated to the complex
phases of the trilinear terms, contribute to the
electric dipole moment of the neutron. For $A_{t}=O(1)$, this
contribution is about $10^{2}-10^{3}$ bigger
than the experimental values \ci{r74}.
Furthermore
an $\alpha\simeq 0.3$ minimizes the unification scale \ci{r59}
and allows for unifying the standard model gauge
groups with the hidden sector one and the fine tuning parameter
$\Delta$ is smallest.  Thus, from all the range of  possible
values for the parameters $n_{g},\,b_{0}$ and $\alpha$,
consistency with the MSSM and electroweak breaking,
selects a very narrow band.
It is remarkable that the values of all soft supersymmetric terms
improve within this band taking quite acceptable values.

\vspace{0.5cm}

\begin{center}
\begin{tabular}{|l|l|l|l|l|l|l|l|l|l|l|l|r|} \hline
$n_{l}$ & $Y$ & $T_{r}$ & $m_{3/2}$ & $\frac{m_{0}}{m_{3/2}}$ &
$\frac{m_{1/2}}{m_{3/2}}$
&$\Lambda_{gut}$ & $\hat\alpha_{gut}^{-1}$ & $A_{t}$ & $B'$ &
$B''$ &
$B'''$ &$\Delta $\\ \hline

3& 4.25& 17.3&55 &4.3& 0.9&2.8&26.7&-0.07&-0.32&18.6&130.5 &
6.7\\ \hline

2&4.22&24.2&82&4.7&0.9&2.8&26.5&-0.07&-0.32&21.8&153& 17.5 \\
\hline

1&4.15&44.5&147&5.3&0.9&2.8&26.1&-0.07&-0.32&26.1&196&71

\\ \hline
\end{tabular}
\end{center}
{\bf Table 1}\hspace{0.3cm}  $SU(6)$ gauge group
with $16\pi^{2}\,b_{0}=15$ and $\alpha_{0}=0.3$.
\vspace{0.5cm}

\begin{center}
\begin{tabular}{|l|l|l|l|l|l|l|l|l|l|l|l|r|} \hline
$n_{l}$ & $Y$ & $T_{r}$ & $m_{3/2}$ & $\frac{m_{0}}{m_{3/2}}$ &
$\frac{m_{1/2}}{m_{3/2}}$
&$\Lambda_{gut}$ & $\hat\alpha_{gut}^{-1}$ & $A_{t}$ & $B'$ &
$B''$ &
$B'''$&$\Delta$ \\ \hline

3& 4.26& 16.3&440 &4.2& 0.9&3.5&26.7&-0.07&-0.32&17.5&122.4
&403\\ \hline

2&4.23&22.5&555&4.4&0.9&3.5&26.6&-0.07&-0.32&19.9&140&720 \\
\hline

1&4.17&40.3&900&5.1&0.9&3.5&25.8&-0.07&-0.32&26.3&184&2508

\\ \hline
\end{tabular}
\end{center}
{\bf Table 2}\hspace{0.3cm}  $SU(6)$ gauge group
with $16\pi^{2}\,b_{0}=16$ and $\alpha_{0}=0.3$.
\vspace{0.5cm}

\begin{center}
\begin{tabular}{|l|l|l|l|l|l|l|l|l|l|l|l|r|} \hline
$n_{l}$ & $Y$ & $T_{r}$ & $m_{3/2}$ & $\frac{m_{0}}{m_{3/2}}$ &
$\frac{m_{1/2}}{m_{3/2}}$
&$\Lambda_{gut}$ & $\hat\alpha_{gut}^{-1}$ & $A_{t}$ & $B'$ &
$B''$ &
$B'''$&$\Delta$ \\ \hline

3&4.62&17.9&632&15.4&0.9&2.0&29.0&-0.8&-1.06&238&158 &$10^{4}$\\
\hline

2&4.61&24.7&661&15.4&0.9&2.0&28.9&-0.8&-1.06&238&158 &$10^{4}$\\
\hline

1&4.57&44.0&849&15.4&0.9&2.0&28.7&-0.8&-1.06&238&158 &$10^{4}$

\\ \hline
\end{tabular}
\end{center}
{\bf Table 3}\hspace{0.3cm}  $SU(6)$ gauge group
with $16\pi^{2}\,b_{0}=11$ and $\alpha_{0}=-1/3$.

\vspace{0.5cm}
\hspace{.5cm} {\small We show in {\bf tables 1-3}, the values for
phenomenologically  relevant parameters determining the
supersymmetry
breaking.
$n_{l}$ is the number of moduli, $T_{r}$, with large v.e.v.,
$g^{-2}_{gut}=Y/2$ is the unification coupling, $m_{3/2}$ is
given in
$GeV$, $\Lambda_{gut}$ in $10^{16}\,GeV$ and $\hat{\alpha}_{gut}$
is
the fine structure constant at $\Lambda_{gut}$. $A_{t}, B', B''$
and
$B''$ are explained in eqs.(\ref{eq:R3}-\ref{eq:R2PP}).  $\Delta$
measures
the tuning of the electroweak breaking (cf. eq.(\ref{eq:D})). The
unification scale is given for $\alpha_{a}=\alpha_{0}=0.3$ in
{\bf tables
1 and 2}, where
$\alpha_{a}$ corresponds to the visible sector gauge groups. With
this
choice the $SU(6)$ and the standard model gauge groups are
unified \ci{r59}.
In {\bf table 3} $\alpha_{a}=0.32$ and the $SU(6)$ gauge group
does not become unified with the visible sector ones.}

\vspace{0.5cm}

To summarise, we have considered the phenomenological
implications of SUSY breaking triggered by a gaugino condensate
in the hidden sector. In a supergravity theory the strong binding
effects play a role in all sectors of the theory and must be
included. Here we have modelled these effects via a simple (N=1
locally supersymmetric) four gaugino interaction which allowed
us to use N-J-L techniques to estimate the nonperturbative
physics. We thus constructed the modular invariant effective
potential which determines the magnitude of the gaugino
condensate and also the moduli of the theory.

We have generalized the work in \ci{r52} by including several
moduli fields. As a result of minimizing the scalar potential we
found that the v.e.v. of the dilaton is mainly given by the
dimension of the hidden sector gauge group. For reasonable hidden
sector gauge groups a large mass hierarchy can arise together
with an acceptable value for the fine structure constant at the
string scale with only one gaugino condensate. We found that the
extremum eqs. for the moduli are solved for a unique value of the
parameter $\alpha$ (cf. eq.(\ref{alpha})) and thus the v.e.v. of
the moduli take a common value ($T_{i}=T_{j}$) related  to
$b_{0},\alpha$ and the v.e.v of the dilaton or they take the dual
invariant values ($T=1,e^{i\pi/6}$). This result naturally
accommodates the squeezed orbifold solutions which are
better candidates for minimal string unification.

We also showed that it is possible to fine tune the cosmological
constant to zero while having SUSY broken at a phenomenologically
interesting value. Incorporating chiral matter fields we computed
the soft SUSY breaking terms.  The ratio between the scalar and
gaugino mass is given  and it depends on $b_{0},\alpha$ and the
v.e.v. of the dilaton.  The numerical values for
phenomenologically interesting solutions are consistent with
results obtained from requiring acceptable values for the gauge
couplings, electroweak breaking, dark matter abundance and
$m_b=m_{\tau}$ assuming a minimal unification of the
MSSM\cite{r58}.

\newpage

\begin{figure}
\vspace{10cm}
\caption{$m_{0}/m_{3/2},\,m_{1/2}/m_{3/2}$ and
the trilinear term $A_{t}$ as a function of $\alpha$. The short-
dashed line
represents $m_{0}/m_{3/2}$, the dot-dashed one $m_{1/2}/m_{3/2}$
while the
long-dashed one $A_{t}$.}
\end{figure}




\begin{thebibliography}{99}
\bibitem{r1}For a review of string theories, see M. Green, J.
Schwarz and E. Witten, Superstring Theory, Cambridge University
Press, 1987.

\bibitem{r34} L.E. Ibanez and D. Lust, Nucl. Phys. B382 (1992)
305.

\bibitem{r73} V. Kaplunovsky and J. Louis, CERN-TH.6809/93;
R. Barbieri, J. Louis and M. Moretti CERN-TH.6856/93.
A.Brignole, L.E.Ibanez and C.Munoz, Universidad Autonoma de
Madrid preprint FTUAM-26/93
\bibitem{r73a}  A. Font, L.E. Ib\'a\~nez, D. L\"ust and F.
Quevedo, Phys. Lett. B245 (1990) 401.
M.Cvetic, A.Font, L.E.Ibanez, D.Lust and F.Quevedo, Nucl.Phys.
B361(1991)194.


\bibitem{r9a} J.P.Derendinger, L.E.Ibanez and H.P.Nilles, Phys.
Lett. B155(1985)65; M.Dine, R.Rohm, N.Seiberg and E.Witten, Phys.
Lett. B156(1985)55.

\bibitem{r9}For a review see D. Amati, K. Konishi, Y. Meurice,
G. Rossi and G. Veneziano, Phys. Rep 162 (1988) 169; H. P.
Nilles, Int. J. Mod. Phys. A5 (1990) 4199 ; J. Louis, SLAC-PUB-
5645 (1991).


 \bibitem{r10} Y. Nambu and G. Jona-Lasinio, Phys. Rev. 122
(1961) 231; D. Gross and A. Neveu, Phys. Rev. D10 (1974) 3235.

\bibitem{r52} A. de la Macorra and G. G. Ross, Nucl. Phys. B404
(1993) 321.

\bibitem{r19} J. P. Derendinger, S. Ferrara, C. Kounas and F.
Zwirner,
Nucl. Phys. B372 (1992) 145.

\bibitem{r18}V. Kaplunovsky, Nucl.Phys. B307 (1988) 145.

\bibitem{r36}L. Dixon, V. Kaplunovsky and J. Louis, Nucl.Phys.
B329
(1990) 27

\bibitem{r15}E. Cremmer, S. Ferrara, L. Girardello and
          A. Van Proeyen, Nucl.Phys. B212 (1983) 413.

\bibitem{r50} G.Veneziano and S. Yankielowicz, Phys. Lett. 113B
(1984);  T.R. Taylor, Phys. Lett 164B (1985) 43
S. Ferrara, N. Magnoli, T.R. Taylor and G. Veneziano, Phys.
Lett. B245 (1990) 409; P. Binetruy and M. K. Gaillard, Phys.
Lett.
232B (1989); Nucl. Phys. B358 (1991) 121.

\bibitem{r55}  H.P. Nilles and M.
           Olechowski, Phys. Lett. B248 (1990) 268; P. Binetruy
and M.K.
           Gaillard, Phys. Lett. B253 (1991) 119; J. Louis,
{\it ``Status of Supersymmetry Breaking in String Theory''},
        SLAC-PUB-5645
           (1991);
            D. L\"ust and T.R. Taylor, Phys. Lett. B253 (1991)
335;
           B. Carlos, J. Casas and C. Mu\~noz, Phys. Lett. B263
(1991) 248;
           S. Kalara, J. Lopez and D. Nanopoulos,
       {\it ``Gauge and Matter Condensates in Realistic String
      Models''},
            preprint CTP-TAMU-69/91.


\bibitem{r17} L. Dixon, V. Kaplunovsky and J. Louis, Nucl.Phys.
B355
(1991) 649; G. Lopes Cardoso and B. Ovrut, Nucl.Phys. B369 (1992)
351;
{\it ``Sigma Model Anomalies, Non-Harmonic Gauge and
Gravitational
 Couplings and String Theory,''} preprint UPR-0481T (1991).

\bibitem{r17a} I. Antoniadis, E. Gava and K.S. Narain, Nucl.Phys.
B383
(1992) 93; Phys.Lett. B283 (1992) 209, I. Antoniadis, K.S. Narain
and
T.R. Taylor, Phys. Lett. B267 (1991) 37.

\bibitem{r56} A. de la Macorra and G.G. Ross, ``{\it
Supersymmetry
Breaking  in 4D String Theory}'',  (preprint OUTP-31P).

\bibitem{clmr}N.V.Krasnikov, Phys. Lett. 193B(1987)37;
L.Dixon, in Proceedings of the APS DPF. Meeting, Houston, 1990;
J.A.Casas, Z.Lalak, C.Munoz, and G.G.Ross, Nucl.Phys.
B347(1990)243;
T.Taylor, Phys.Lett B252(1990)59;
B. de Carlos, J. A. Casas and C. Munoz,
CERN-TH.6436/92
and ref. therein.

 \bibitem{r23} S. Coleman and E. Weinberg, Phys. Rev. D7 (1973)
1883.

\bibitem{r57} R. Barbieri and S. Cecotti, Z. Phys. C 17, (1983)
183;
M. Srednicki and S. Theisen, Phys. Rev. Lett. 54, (1985) 278; P.
Binetruy, S. Dawson, M.K. Gaillard and I. Hinchliffe, Phys. Rev.
D
37 (1988) 2633.

\bibitem{r58}
G. Costa, J. Ellis, G.L. Fogli, D.V. Nanopolous and F. Zwirner,
Nucl.Phys. B297
   (1988) 244;
J. Ellis, S. Kelley and D.V. Nanopoulos, Phys. Lett. B249 (1990)
441;
Phys. Lett. B260 (1991) 131;
P. Langacker, {\it ``Precision tests of the standard model",}
Pennsylvania preprint UPR-0435T, (1990);
U. Amaldi, W. de Boer and H. F\"urstenau, Phys. Lett. B260 (1991)
447;
P. Langacker and M. Luo, Phys.Rev.{\bf D44} (1991) 817;
R. Roberts and G.G. Ross, preprint RAL-92-005 (1992).


\bibitem{r26} A. Font, L.E. Ib\'a\~nez, F. Quevedo and
A. Sierra, Nucl.Phys. B331 (1990)  421.


\bibitem{r54} L.E. Ib\'a\~nez, D. L\"ust and G.G. Ross,
Phys. Lett. B272 (1991) 251.

\bibitem{r67} G. F. Giudice and A. Masiero, Phys. Lett. 206B
(1988) 480;
J.E.~Kim and H.-P.~Nilles, Phys. Lett. B 138B (1984) 150;
J.L.~Lopez and D.V.~Nanopoulos, Phys. Lett. B B251 (1990) 73;
E.J.~Chun,  J.E.~Kim and H.-P.~Nilles, Nucl. Phys. B370 (1992)
105.

\bibitem{r65} R. Barbieri and G.F. Giudice, Nucl. Phys. B302
(1988) 63.

\bibitem{r63} K. Inoue et al., Prog.Theor.Phys. {\bf 68} (1982)
927;
L.E. Ib\'a\~nez, Nucl.Phys. B218 (1983) 514;
L.E. Ib\'a\~nez and C. L\'opez, Phys. Lett. B126 (1983) 54;
Nucl.Phys. B233 (1984) 511;
L. Alvarez-Gaume, J. Polchinsky and M. Wise, Nucl.Phys. B221
(1983) 495; L.E Ibanez, C. Lopez and C. Munoz, Nucl. Phys. B256
(1985) 218.


\bibitem{r74} J.E. Kim, Phys. Rep. 150 (1987)1; H.Y. Cheng, Phys.
Rep.
158 (1988) 1; L. E. Ibanez and D. Lust, Nucl.
Phys. B267 (1991) 51.

\bibitem{r59} A. de la Macorra, ``{\it Unification scale
in string theory}'', (preprint OUTP-93-33P).

\end{thebibliography}
\end{document}